\documentclass[aps,prl,twocolumn,superscriptaddress,citesort]{revtex4-1} 

\usepackage{graphicx}
\usepackage{amsmath}
\usepackage{amssymb}
\usepackage{epstopdf}
\usepackage{color}
\usepackage{float}
\usepackage{bm}
\def\build#1_#2^#3{\mathrel{\mathop{\kern 0pt#1}\limits_{#2}^{#3}}}

\usepackage{soul}

\begin{document}
		
		\title{Flow of Spatiotemporal Turbulentlike Random  Fields}
		
		
		
		\author{Jason Reneuve}
		\email[]{jason.reneuve@ens-lyon.fr}
		\affiliation{Laboratoire des Ecoulements G\'eophysiques et Industriels, Universit\'e Grenoble Alpes, CNRS, Grenoble-INP, F-38000 Grenoble, France}			

		\author{Laurent Chevillard}
		\affiliation{Univ Lyon, Ens de Lyon, Univ Claude Bernard, CNRS, Laboratoire de Physique, 46 all\'ee d'Italie F-69342 Lyon, France}

		\date{\today}
		
		\begin{abstract}
			We study the Lagrangian trajectories of statistically isotropic, homogeneous, and stationary divergence free spatiotemporal random vector fields. We design this advecting Eulerian velocity field such that it gets asymptotically rough and multifractal, both in space and time, as it is demanded by the phenomenology of turbulence at infinite Reynolds numbers. We then solve numerically the flow equations for a  differentiable version of this field. We observe that trajectories get also rough, characterized by nearly the same Hurst exponent as the one of our prescribed advecting field. Moreover, even when considering the simplest situation of the advection by a fractional Gaussian field, we evidence in the Lagrangian framework additional intermittent corrections. The present approach involves properly defined random fields, and asks for a rigorous treatment that would explain our numerical findings and deepen our understanding of this long lasting problem. 
			
		\end{abstract}
	\maketitle
	

	A powerful and physically insightful way to characterize many dynamical systems, such as those encountered in fluid mechanics, consists in studying the path lines $\boldsymbol{X}(t)$ of a given advecting field $\boldsymbol{u}(\boldsymbol{x},t)$, at the position $\boldsymbol{x}\in \mathbb R^d$ and time $t>0$, defined by
	\begin{equation}
		\label{eq:FlowLines}
		\frac{d\boldsymbol{X}(t)}{dt}=\boldsymbol{u}(\boldsymbol{X}(t),t).
	\end{equation}
	In the context of fluid turbulence, where the velocity field $\boldsymbol{u}$ is governed by the Navier-Stokes equations, such Lagrangian trajectories of fluid particles have been extensively studied in laboratory and numerical flows  \cite{YeuPop89,VotSat98,PorVot01,MorMet01,MorDel02,MorDel03,CheRou03,Fri03,BifBof04,BecBif06,TosBod09,PinSaw12,BenLal19}. In this situation, the three-dimensional Eulerian advecting flow $\boldsymbol{u}$ is incompressible (i.e. divergence free) and exhibits a complex multiscale structure in both space \cite{Fri95} and time \cite{TenLum72}. In particular, in the fully developed turbulent regime concerning the asymptotic limit of infinite Reynolds numbers, $\boldsymbol{u}$ gets rough (i.e. nondifferentiable) in both space and time, and characterized in a statistically averaged sense by a Hurst exponent  of order $H_{\text{\tiny{Eul}}}\approx 1/3$. In the phenomenology of turbulence mostly developed by Kolmogorov \cite{Kol41}, this can be broadly understood on dimensional grounds if it is assumed that the average dissipation by unit of mass remains finite at infinite Reynolds numbers \cite{Fri95}. Similarly, the Lagrangian velocity $\boldsymbol{v}(t)\equiv \boldsymbol{u}(\boldsymbol{X}(t),t)$, i.e. the velocity of a tracer advected by the flow $\boldsymbol{u}$, develops small scales such that it gets rough and characterized by a Hurst exponent  of order $H_{\text{\tiny{Lag}}}\approx 1/2$. Again, under the same assumption, this exponent can be obtained from dimensional arguments, and says that Lagrangian velocity has the same regularity as the one of a Brownian motion \cite{TenLum72}. 
	
	Whereas it remains elusive to derive these behaviors from first principles, we propose in this Letter to study the statistical properties of Lagrangian trajectories extracted from a prescribed advecting velocity field that reproduces some of the main aforementioned features of turbulence. A similar approach has been already explored for various random vector fields \cite{KomPap97,FunVas98,FanKom00,ChaGaw03,KhaVas04}, although, as we will see, our advecting flow is more general, in particular concerning possible intermittent corrections. 

	In order to draw the simplest and numerically tractable picture of these phenomena, we need to come up with a proposition for the prescribed advecting velocity field $\boldsymbol{u}(\boldsymbol{x},t)$. Recall that we want it to be divergence free at any time to ensure statistical stationarity of induced Lagrangian velocities \cite{FalGaw01}. For this reason, we will consider henceforth a two-component vector field $\boldsymbol{u}=(u_1,u_2)$ living in a two-dimensional space $\boldsymbol{x}=(x_1,x_2)\in\mathbb R^2$ and for $t\in\mathbb{R}$, such that $\boldsymbol{\nabla}\cdot\boldsymbol{u}=0$ at any time. In an asymptotic regime, mimicking the behavior of turbulence at infinite Reynolds numbers, this vector field is eventually rough, governed in a statistically averaged sense by a Hurst exponent $H\in ]0,1[$ (taken to be $1/3$ as far as turbulence is concerned). A first step in this direction would be to consider fractional Gaussian fields, defined as linear operations on a space-time white noise (similarly to the approach developed in \cite{RobVar08,CheRob10,PerGar16,CheGar19,Ren19}), regularized over a small parameter $\epsilon>0$ ensuring differentiability in both space and time (compatible in particular with the divergence free condition). Going beyond this Gaussian framework, we would like also to consider some intermittent (i.e. multifractal) corrections \cite{Fri95}, and to explore their implication on the statistical behavior of Lagrangian trajectories. To make our notations lighter, without loss of generality, we consider in the sequel nondimensional space and time coordinates. 
	
	Along these lines, the simplest random vector field that we have in mind, which is statistically stationary, isotropic and homogeneous, and which reproduces these statistical behaviors, is given by
	\begin{equation}\label{eq:VelField}
		\boldsymbol{u}(\boldsymbol{x},t) =  \int_{\boldsymbol{y}\in\mathbb R^2, s\in\mathbb R}\boldsymbol{\mathcal G}_{\epsilon,H_{\text{\tiny{Eul}}}}(\boldsymbol{x}-\boldsymbol{y},t-s)M_{\epsilon,\gamma_{\text{\tiny{Eul}}}}(d^2y,ds),
	\end{equation}
	where the vector kernel $\boldsymbol{\mathcal G}_{\epsilon,H_{\text{\tiny{Eul}}}}$ acting linearly on the random measure $M_{\epsilon,\gamma_{\text{\tiny{Eul}}}}$ (specified later) reads
	\begin{equation}\label{eq:KernelG}
		\boldsymbol{\mathcal G}_{\epsilon,H_{\text{\tiny{Eul}}}}(\boldsymbol{x},t)=\varphi(\boldsymbol{x},t)\frac{\boldsymbol{x}^\perp}{||\boldsymbol{x},0||_\epsilon}||\boldsymbol{x},t||_\epsilon^{H_{\text{\tiny{Eul}}}-3/2},
	\end{equation}
	with $||\boldsymbol{x},t||_\epsilon^2 = |\boldsymbol{x}|^2+t^2+\epsilon^2$ a regularized spatiotemporal norm over $\epsilon$ and $\boldsymbol{x}^\perp=(-x_2,x_1)$. Note that we implicitly assume that in our nondimensional reference frame, the small scale $\epsilon$ plays the role of both the spatial and temporal dissipative scales. This is consistent with the similar dependence of the so-called Kolmogorov length scale $\eta_K$ and the sweeping timescale \cite{TenLum72} on the Reynolds number. The scalar cutoff function $\varphi$ ensures that this field has a finite variance,. It goes smoothly to zero as $|\boldsymbol{x}|$ gets of the order of the integral length scale $L$ and/or $t$ of the order of the integral timescale $T$. Once expressed in our nondimensional coordinate system, we take $L=T$ and assume $\varphi(\boldsymbol{x},t)=\exp\left(-\frac{|\boldsymbol{x}|^2+t^2}{2L^2}\right)$. The very form of the kernel $\boldsymbol{\mathcal G}$ (Eq. (\ref{eq:KernelG})) is inspired by the two-dimensional Biot-Savart law \cite{MajBer02}, and ensures that the velocity field (Eq. (\ref{eq:VelField})) is divergence free for any finite $\epsilon>0$ and at any time. Additional technical details are provided in \cite{SM}.

	The random spatiotemporal measure $M_{\epsilon,\gamma_{\text{\tiny{Eul}}}}$ reads
	\begin{equation}\label{eq:MGamma}
		M_{\epsilon,\gamma_{\text{\tiny{Eul}}}}(d^2y,ds)=e^{\gamma_{\text{\tiny{Eul}}} Y_\epsilon(\boldsymbol{y},s)-\gamma_{\text{\tiny{Eul}}}^2\langle Y_\epsilon^2\rangle}W(d^2y,ds),
	\end{equation}
	where  $W$ is a spatiotemporal Gaussian white noise (thus $2+1$-dimensional) and $Y_\epsilon$ a zero-average scalar Gaussian random field, logarithmically correlated in both space and time as $\epsilon\rightarrow 0$, taken as independent of $W$. As we will see, the parameter $\gamma_{\text{\tiny{Eul}}}$ governs entirely the intermittent corrections, and $M_{\epsilon,\gamma_{\text{\tiny{Eul}}}}$ can be viewed as a continuous, statistically homogeneous and stationary version of the discrete cascade models \cite{MenSre87,BenBif93,ArnBac98}. Being Gaussian, the scalar field $Y_\epsilon$ can be obtained as a linear operation on an independent white noise $\widetilde{W}$, that is $Y_\epsilon(\boldsymbol{x},t) = \frac{1}{\sqrt{4\pi}}\int_{\boldsymbol{y},s}\mathcal H_\epsilon(\boldsymbol{x}-\boldsymbol{y},t-s)\widetilde{W}(d^2y,ds)$ with $\mathcal H_\epsilon(\boldsymbol{x},t) = ||\boldsymbol{x},t||_\epsilon^{-3/2}1_{|\boldsymbol{x}|^2+t^2\le L^2}$ and $1_S$ the indicator function of the set $S$.
	
	Using similar technics as in Refs. \cite{RobVar08,CheRob10,PerGar16,CheGar19,Ren19}, in particular calling for stochastic calculus methods developed for multiplicative chaos theory \cite{RhoVar14}, it can be shown that the velocity field $\boldsymbol{u}$ (Eq. (\ref{eq:VelField})) is rough in the limit of vanishing regularizing scale $\epsilon\rightarrow 0$, such that for instance the moments of the longitudinal velocity increments $\delta_\ell u_1(\boldsymbol{x},t)=u_1(x_1+\ell,x_2,t)-u_1(x_1,x_2,t)$ (i.e. the structure functions) behave for $q\ge 1$, $H_{\text{\tiny{Eul}}}\in]0,1[$ and $\gamma^2\le H_{\text{\tiny{Eul}}}/(q-1)$, as
	\begin{equation}\label{eq:SFEul}
		\lim_{\epsilon\rightarrow 0}\langle(\delta_\ell u_1)^{2q}\rangle\build{\sim}_{\ell\rightarrow 0^+}^{}C_{2q,H_{\text{\tiny{Eul}}},\gamma_{\text{\tiny{Eul}}}}\ell^{2qH_{\text{\tiny{Eul}}}-2q(q-1)\gamma_{\text{\tiny{Eul}}}^2},
	\end{equation}
	where the multiplicative factor $C_{2q,H,\gamma_{\text{\tiny{Eul}}}}$ is finite and positive. The scaling behavior entering in Eq. (\ref{eq:SFEul}) indicates that $\boldsymbol{u}$ (Eq. (\ref{eq:VelField})) is intermittent and exhibits a quadratic (i.e. log-normal) spectrum. The respective transverse (i.e. the scale $\ell$ is taken along the second direction) and temporal (i.e. we look at the increment over a time $\tau$ at a fixed position) structure functions behave similarly as in Eq. (\ref{eq:SFEul}), with the same spectrum of exponents but with different multiplicative constants. More general spectra than the quadratic one could be considered \cite{BarMan02,SchMar01,BacMuz03,RhoVar14}, although calculations leading to the exact asymptotic result  Eq. (\ref{eq:SFEul}) get more intricate, and the quadratic spectrum reproduces a convincing phenomenology of intermittency at low statistical orders.

		\begin{figure}
		\begin{center}
			\includegraphics[width=1\columnwidth]{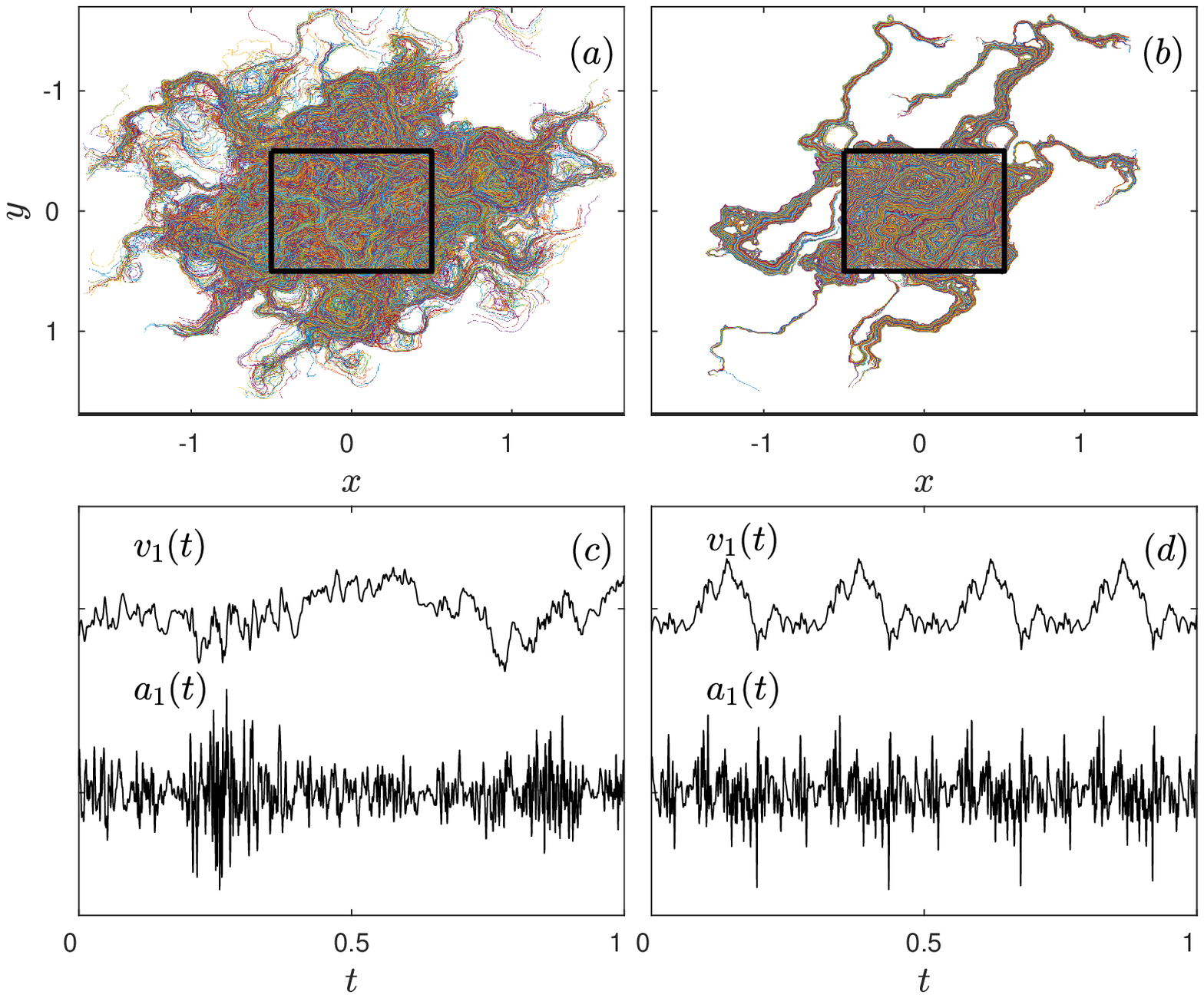}
			\vspace{-0.2cm}
			\caption{Path lines $\boldsymbol{X}(t)$ (Eq. (\ref{eq:FlowLines})) of a Gaussian velocity field $\boldsymbol{u}(\boldsymbol{x},t)$ (Eq. (\ref{eq:VelField})), using $H_{\text{\tiny{Eul}}}=1/3$ and $\gamma_{\text{\tiny{Eul}}}=0$. Other parameters of the simulation are given in the text. (a) Each trajectories are represented with various colors, starting initially from positions uniformly distributed in the unit square centered on the origin (and represented with thick black lines). (c) Typical time series of velocity $v_1$ and acceleration $a_1$ of a particle. Series are arbitrarily shifted horizontally and renormalized such that they are of same variance. (b) and (d) Similar plot as in (a) and (c), but for a frozen-in-time velocity field $\boldsymbol{u}(\boldsymbol{x},0)$. }
			\label{fig:1}				
		\end{center}
		\vspace{-0.7cm}
	\end{figure}

Numerical simulations of $\boldsymbol{u}$ (Eq. (\ref{eq:VelField})) are performed in a $(2+1)$-dimensional periodic box of unit length and duration using $N=2^{11}$ collocations points in each direction, such that $dx=dt=1/N$. Convolutions of the deterministic functions $\boldsymbol{\mathcal G}_{\epsilon,H}$ (Eq. (\ref{eq:KernelG})) and $\mathcal H_\epsilon$ (i.e. the kernel of $Y_\epsilon$ entering in Eq. (\ref{eq:MGamma})) with two independent instances $W$ and $\widetilde{W}$ of variance $dx^2dt$ of the white noise are computed in an efficient way in the Fourier domain. We use for the large scales $L=T=1/4$. The singular kernels $\boldsymbol{\mathcal G}_{\epsilon,H}$ and $\mathcal H_\epsilon$ are regularized over the small scale $\epsilon=4dx$ such that, up to numerical errors, the obtained field $\boldsymbol{u}$ is differentiable in space and time, and divergence free in particular. Finally, the trajectories $\boldsymbol{X}(t)$ of $2^{14}$ particles, initially uniformly distributed in the unit square, are computed according to Eq. (\ref{eq:FlowLines}) using a second-order Runge-Kutta time marching scheme and linear interpolation of the velocities, as detailed in Ref. \cite{YuKan12}. Their respective Lagrangian velocity $\boldsymbol{v}(t)=d\boldsymbol{X}(t)/dt$ and acceleration $\boldsymbol{a}(t)=d^2\boldsymbol{X}(t)/dt^2$ are obtained using finite-difference time derivatives.

	Let us first focus on the statistical analysis of the trajectories in an advecting Gaussian velocity field $\boldsymbol{u}(\boldsymbol{x},t)$ (Eq. (\ref{eq:VelField})). To do so, we consider the nonintermittent case $\gamma_{\text{\tiny{Eul}}}=0$, and the particular value $H_{\text{\tiny{Eul}}}=1/3$ to mimic the regularity of turbulence. We display in Fig. \ref{fig:1}(a) the trajectories of particles initially uniformly distributed in the unit square. We indeed observe strong chaotic mixing, and notice that during the unit duration of the simulation, particles have traveled a distance of order unity, as expected. We show in Fig. \ref{fig:1}(c) typical time series of velocity $v_1(t)$ and acceleration $a_1(t)$ over the duration of the simulation. We can see that series are indeed statistically stationary. Also, $v_1$ is correlated over the large integral timescale $T$, whereas $a_1$ gets correlated over the small timescale $\epsilon$, which is consistent with the phenomenology of turbulence. A trained eye would see that $a$ clearly deviates from Gaussianity. 
	
	At this stage, it is tempting to explore the statistics of the trajectories obtained while advecting the tracers by a frozen-in-time velocity field, say $\boldsymbol{u}(\boldsymbol{x},0)$. We represent in Fig. \ref{fig:1}(b) the respective trajectories. Mixing is there much less efficient than for the time evolving velocity field (Fig. \ref{fig:1}(a)). In particular, many of them have closed orbits. Typical time series of $v_1$ and $a_1$ on a closed orbit are shown in Fig. \ref{fig:1}(d), displaying an expected periodicity.

	\begin{figure}
		\begin{center}
			\includegraphics[width=1\columnwidth]{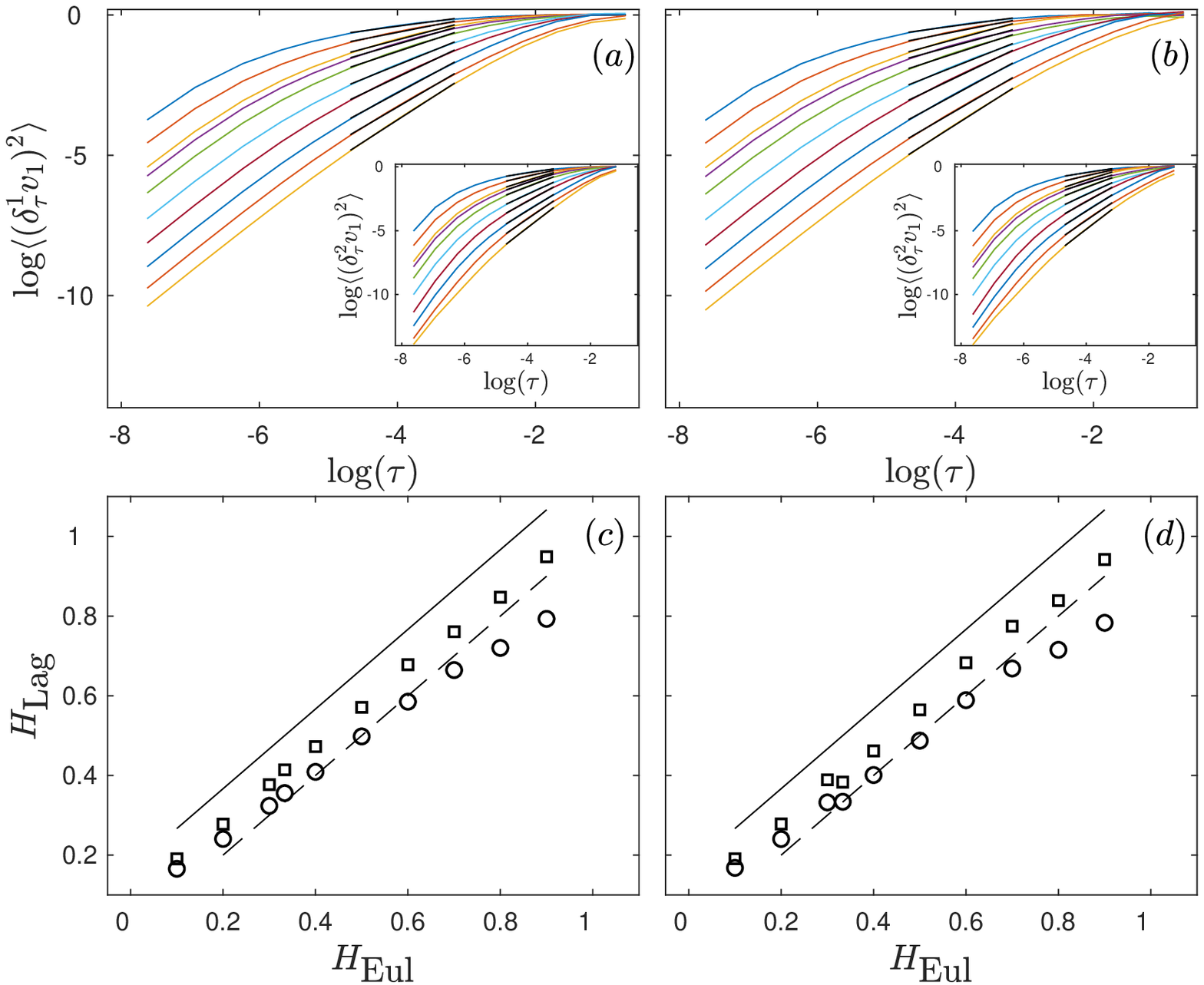}
			\vspace{-0.2cm}
			\caption{(a) Logarithmic representation of the second-order Lagrangian structure function $\langle(\delta_\tau^1 v_1)^{2}\rangle/(2\langle v_1^{2}\rangle)$ (Eq. (\ref{eq:SFLag})) obtained from a Gaussian velocity field $\boldsymbol{u}(\boldsymbol{x},t)$ (Eq. (\ref{eq:VelField})) using $\gamma_{\text{\tiny{Eul}}}=0$ and   $H_{\text{\tiny{Eul}}}=0.1, 0.2, 0.3, 1/3, 0.4, 0.5, 0.6, 0.7, 0.8$ and $0.9$ (from top to bottom). Results of our fitting procedure are displayed with black lines. Inset: Similar plot as in (a), but for the second-order velocity increment moment $\langle(\delta_\tau^2 v_1)^{2}\rangle/(6\langle v_1^{2}\rangle)$. (b) Same plot as in (a), but for the frozen-in-time advecting field $\boldsymbol{u}(\boldsymbol{x},0)$. (c) Power-law exponents observed in (a), i.e. $2H_{\text{\tiny{Lag}}}$, estimated using $\langle(\delta_\tau^1 v_1)^{2}\rangle$ ($\circ$) and $\langle(\delta_\tau^2 v_1)^{2}\rangle$ ($\square$). We superimpose the two discussed behaviors $H_{\text{\tiny{Lag}}}=H_{\text{\tiny{Eul}}}$ (dashed line) and $H_{\text{\tiny{Lag}}}=H_{\text{\tiny{Eul}}} + \frac{1}{6}$ (solid line). (d) Same plot as in (c), but for $\boldsymbol{u}(\boldsymbol{x},0)$.} 
			\label{fig:2}				
		\end{center}
		\vspace{-0.7cm}
	\end{figure}

	Let us now estimate the regularity of $\boldsymbol{v}(t)$ obtained from a Gaussian velocity field $\boldsymbol{u}(\boldsymbol{x},t)$ (Eq. (\ref{eq:VelField}) with $\gamma_{\text{\tiny{Eul}}}=0$), and quantify its dependence on $H_{\text{\tiny{Eul}}}$. To do so, we perform simulations using ten values for $H_{\text{\tiny{Eul}}}$ between $0.1$ and $0.9$. Subsequent statistics are obtained using $2^{14}$ trajectories from ten independent realizations of the random Eulerian field. To quantify the regularity of $\boldsymbol{v}$, we estimate the moments of the velocity time increments $\delta_\tau^1 v_1(t)=v_1(t+\tau)-v_1(t)$, and define the respective Lagrangian Hurst exponent $H_{\text{\tiny{Lag}}}$ and intermittency coefficient $\gamma_{\text{\tiny{Lag}}}$ as
	\begin{equation}\label{eq:SFLag}
		\langle(\delta_\tau^1 v_1)^{2q} \rangle \build{\propto}_{\epsilon\ll \tau\ll T}^{}\tau^{2qH_{\text{\tiny{Lag}}}-2q(q-1)\gamma^2_{\text{\tiny{Lag}}}},
	\end{equation}
	such that $H_{\text{\tiny{Lag}}}$ can be estimated while fitting in the inertial range (i.e. for $\epsilon\ll \tau\ll T$) the power-law exponent of $\langle(\delta_\tau^1 v_1)^{2}\rangle\propto \tau^{2H_{\text{\tiny{Lag}}}}$. More generally, let us note that whereas the behavior of Eulerian structure functions (Eq. (\ref{eq:SFEul})) is exact in the asymptotic limit of vanishing $\epsilon$ and scale $\ell$, the proposed behavior of their Lagrangian counterparts (Eq. (\ref{eq:SFLag})) is a model whose parameters $(H_{\text{\tiny{Lag}}},\gamma^2_{\text{\tiny{Lag}}})$ will be eventually estimated following a fitting procedure. 

	We display in Fig. \ref{fig:2}(a) the dependence on the scale $\tau$ of the second-order structure function in a logarithmic representation, for the ten values of the Eulerian Hurst exponent $H_{\text{\tiny{Eul}}}$. We indeed observe a power-law behavior between the dissipative range $\tau\ll\epsilon$, where $\langle(\delta_\tau^1 v_1)^{2}\rangle\propto \tau^2$ and the large scales $\tau\gg T$ for which we get a saturation toward $2\langle v_1^{2}\rangle$. We proceed with the fit of the power-law exponent (represented by solid black lines) and gather our results in Fig. \ref{fig:2}(c) (using $\circ$). We can see that the estimated regularity of Lagrangian trajectories $H_{\text{\tiny{Lag}}}$ is very close to the imposed Eulerian regularity $H_{\text{\tiny{Eul}}}$, that is $H_{\text{\tiny{Lag}}}\approx H_{\text{\tiny{Eul}}}$, as it was observed in the synthetic three-dimensional, slowly evolving in time, flow of Ref. \cite{KhaVas04} and in the frozen Navier-Stokes field of Ref. \cite{CheRou05}. We superimpose with a dashed line such a prediction, showing that is does reproduce some of our estimations when $H_{\text{\tiny{Eul}}}$ is smaller than $1/2$. Since the level of regularity is high, it is tempting to check whether similar results are obtained with the second-order increment, that is the increments of the increments $\delta_\tau^2 v_1(t)=\delta_\tau^1 v_1(t+\tau)-\delta_\tau^1 v_1(t)$, which is not only orthogonal to constants, but also to local linear trends, allowing in particular to estimate Hurst exponents greater than unity. We display in the inset of Fig.  \ref{fig:2}(a) the behavior of their second moment as a function of the scale $\tau$. Once again, we observe a power-law behavior between the dissipative range, where $\langle(\delta_\tau^2 v_1)^{2}\rangle\propto \tau^4$ and the large scales $\tau\gg T$ for which we get a saturation toward $6\langle v_1^{2}\rangle$. We fit the obtained exponents and reproduce our results in Fig. \ref{fig:2}(c) (using $\square$). In this case, we obtain a very convincing linear behavior, that falls in between $H_{\text{\tiny{Eul}}}$ (dashed line) and  $H_{\text{\tiny{Lag}}}=H_{\text{\tiny{Eul}}} + \frac{1}{6}$, that includes in particular the Kolmogorov's values $H_{\text{\tiny{Eul}}}=1/3$ and $H_{\text{\tiny{Lag}}}=1/2$ (represented by a solid black line). We performed the same analysis using the third-order increment, i.e. $\delta_\tau^3 v_1(t)=\delta_\tau^2 v_1(t+\tau)-\delta_\tau^2 v_1(t)$, and obtain same results as with $\delta_\tau^2 v_1$ (data not shown). We report in Figs. \ref{fig:2}(b) and (d) a similar study, but with a frozen-in-time advection velocity field $\boldsymbol{u}(\boldsymbol{x},0)$, as it is illustrated in \ref{fig:1}(b) and (d). The very same conclusions as in the time-evolving case can be drawn. In \cite{SM}, we perform additional numerical simulations, using larger resolutions up to $N=2^{16}$ collocation points of purely spatial advecting fields, that allow to unambiguously eliminate the effects of regularization at small $\epsilon$ and large $L$ scales, which confirm that $H_{\text{\tiny{Lag}}} \approx H_{\text{\tiny{Eul}}}$.

	\begin{figure}
		\begin{center}
			\includegraphics[width=1\columnwidth]{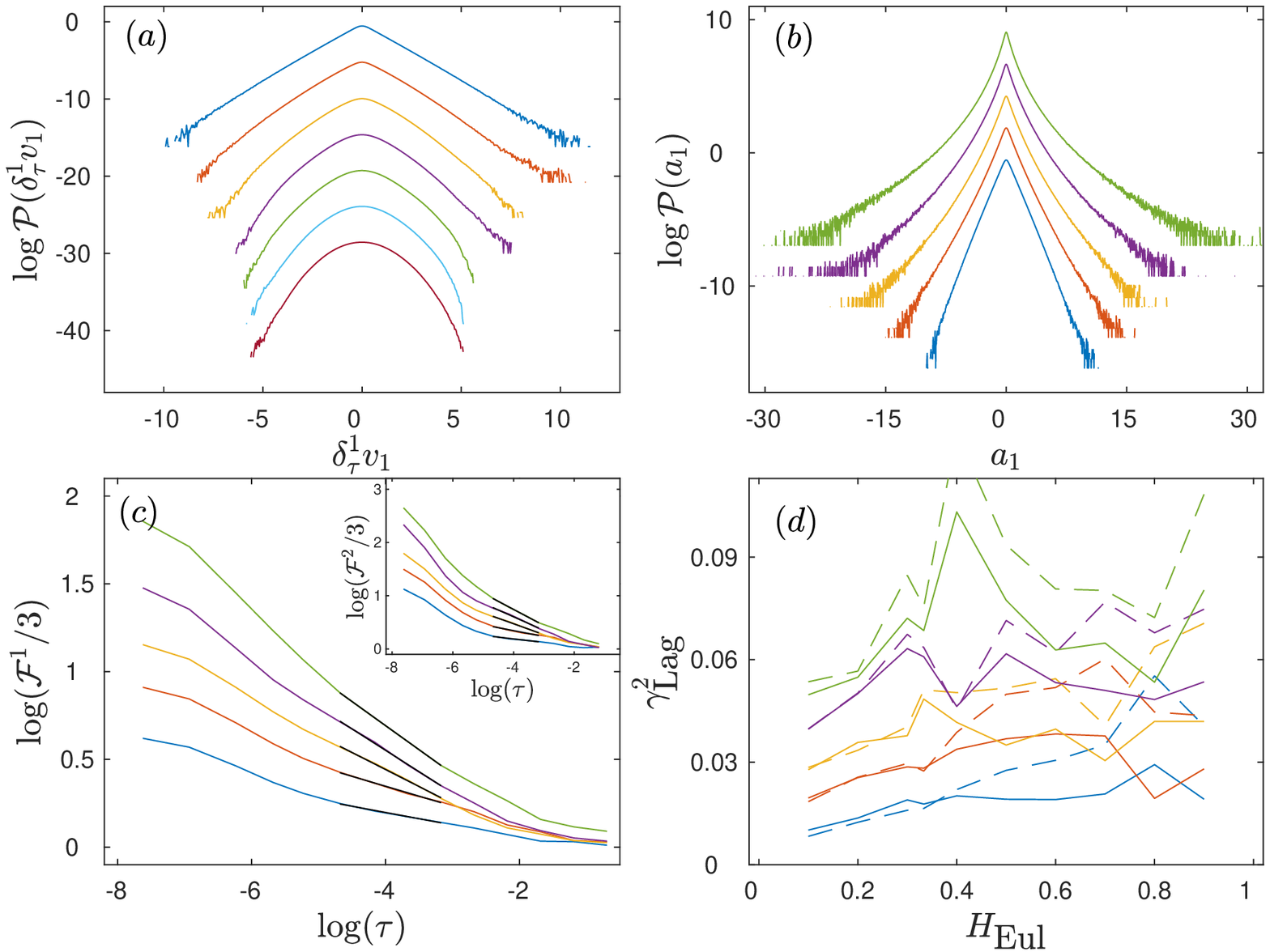}
			\vspace{-0.2cm}
			\caption{(a) PDFs of the Lagrangian increments $\delta_\tau^1 v_1$ from large (bottom) to small (top) scales in a Gaussian advecting field of parameters $H_{\text{\tiny{Eul}}}=1/3$ and $\gamma_{\text{\tiny{Eul}}}=0$. PDFs are all of unit variance, and arbitrarily shifted vertically for clarity. (b) PDFs of Lagrangian acceleration for $H_{\text{\tiny{Eul}}}=1/3$ and for $\gamma^2_{\text{\tiny{Eul}}}=0, 0.01, 0.02, 0.03$ and $0.04$ (from bottom to top), of unit variance and arbitrarily shifted. (c) Logarithmic representation of the flatness of $\delta_\tau^1 v_1$ (see text), with same parameters and colors as in (b). Inset: same as in (c), but for $\delta_\tau^2 v_1$. Results of fitting are displayed with black lines. (d) Estimated values for $\gamma_{\text{\tiny{Lag}}}$ (Eq. (\ref{eq:SFLag})) from the fitting procedure of the flatness curves of (c). Same colors as in (b) and (c), for $\delta_\tau^1 v_1$ (solid lines) and   $\delta_\tau^2 v_1$ (dashed lines).}
			\label{fig:3}				
		\end{center}
		\vspace{-0.7cm}
	\end{figure}

	Let us finally quantify intermittent corrections on the trajectories (i.e. the dependence of $H_{\text{\tiny{Lag}}}$ and $\gamma_{\text{\tiny{Lag}}}$ on $H_{\text{\tiny{Eul}}}$ and $\gamma_{\text{\tiny{Eul}}}$). To do so, we repeat former simulations for five values of the parameter $\gamma_{\text{\tiny{Eul}}}$. Recall that in a 3$d$ turbulent field,  $\gamma^2_{\text{\tiny{Eul}}}\approx 0.025$ \cite{Fri95}.  We start by performing a similar study as presented in Fig. \ref{fig:2}, but with a varying $\gamma_{\text{\tiny{Eul}}}$, and found no differences with former conclusions: $H_{\text{\tiny{Lag}}}\approx H_{\text{\tiny{Eul}}}$, independently of $\gamma_{\text{\tiny{Eul}}}$ (data not shown). This is a nontrivial property. Furthermore, trajectories extracted from a Gaussian field (i.e. $\gamma_{\text{\tiny{Eul}}}=0$) are intermittent. To see this, we display in Fig. \ref{fig:3}(a) the Probability Density Functions (PDFs) of Lagrangian velocity increments at various scales (using $H_{\text{\tiny{Eul}}}=1/3$ and $\gamma_{\text{\tiny{Eul}}}=0$). We indeed observe the continuous shape deformation of the PDFs, which is characteristic of the intermittency phenomenon \cite{CasGag90}. Actually, these non-Gaussian behaviors were already seen on the typical time series of acceleration in Fig. \ref{fig:1}(c). Note that at the smallest scale (top blue curve of Fig. \ref{fig:3}(a)), PDFs of increments and acceleration coincide in this representation, and exhibit noticeable exponential tails, as they are obtained for pressure gradients in Gaussian ensembles \cite{HolSig93}. In the same line, we represent in Fig. \ref{fig:3}(b) the acceleration PDFs for varying $\gamma_{\text{\tiny{Eul}}}$, and for $H_{\text{\tiny{Eul}}}=1/3$. We see that as $\gamma_{\text{\tiny{Eul}}}$ increases, the acceleration PDF develops larger and larger tails, which shows that $\gamma_{\text{\tiny{Lag}}}$ increases in a monotonic way with $\gamma_{\text{\tiny{Eul}}}$. To quantify more precisely this dependence, we estimate the Lagrangian velocity Flatness $\mathcal F^1(\tau)=\langle(\delta_\tau^1 v_1)^{4}\rangle/\langle(\delta_\tau^1 v_1)^{2}\rangle^2$ that is expected to behave, according to Eq. (\ref{eq:SFLag}), as $\tau^{-4\gamma^2_{\text{\tiny{Lag}}}}$ in the inertial range. We represent in Fig. \ref{fig:3}(c) the behavior of the flatness for increasing values of $\gamma_{\text{\tiny{Eul}}}$ and $H_{\text{\tiny{Eul}}}=1/3$. We see that flatness is close to three at large scales $\tau\sim T$, i.e. the value for a Gaussian process, and increases, all the more as $\gamma_{\text{\tiny{Eul}}}$ gets bigger, as the scale decreases. The overall dependence of  $\gamma_{\text{\tiny{Lag}}}$ on both  $H_{\text{\tiny{Eul}}}$ and  $\gamma_{\text{\tiny{Eul}}}$ is illustrated in  Fig. \ref{fig:3}(d), where the estimation of $\gamma_{\text{\tiny{Lag}}}$ is based on both the flatness of the first-order (solid lines) and second-order (dashed lines) increments. We can conclude to a complex dependence of $\gamma_{\text{\tiny{Lag}}}$ on the parameters of the advecting Eulerian field. Interestingly, the Lagrangian intermittency coefficient in experimental and numerical 3$d$ flows has been found compatible with $\gamma^2_{\text{\tiny{Lag}}}\approx 0.085$ \cite{CheRou03}, a value which is of the order of what is found presently when we focus on the particular value $H_{\text{\tiny{Eul}}}\approx 1/3$ and $\gamma^2_{\text{\tiny{Eul}}}=0.025$. We provide in \cite{SM} a similar study with a frozen-in-time advecting field that shows that obtained intermittent corrections on Lagrangian velocities are similar to those displayed in Fig. \ref{fig:3}.

	To summarize, we have built an incompressible statistically homogeneous, isotropic and stationary spatiotemporal Eulerian advection field (Eq. (\ref{eq:VelField})). It is asymptotically rough and multifractal (Eq. (\ref{eq:SFEul})), governed at small scales by the parameters $H_{\text{\tiny{Eul}}}$ and $\gamma_{\text{\tiny{Eul}}}$. We have then estimated,  based on numerical simulations, the statistical properties of its Lagrangian trajectories. We find that they are also asymptotically rough and multifractal (Eq. (\ref{eq:SFLag})), and relate their parameters $H_{\text{\tiny{Lag}}}$ and $\gamma_{\text{\tiny{Lag}}}$ to those of the advecting Eulerian field. In particular, we estimate with good accuracy, at second-order from a statistical point of view, that the regularity of the trajectories follows closely the one of the Eulerian field. Furthermore, we evidence unambiguous intermittent corrections, even when the advecting field is prescribed to be Gaussian. These are new and nontrivial results that are calling for new theoretical developments. In this regard, great progress has been made in the mathematical description of path lines of some rough advecting fields \cite{Dub09,MilShe16}. Also, the proposed velocity field could be used to investigate related important situations, such as the passive advection of scalars \cite{FalGaw01}, and the relative dispersion of particle pairs \cite{BouOue06,BitHom12}. Advecting fields of Ref. \cite{ChaGaw03}, some of which get rid of the sweeping by large scales, are explored in \cite{SM}, and lead for some aspects to similar conclusions.  Finally, including the intrinsically asymmetrical  nature of the distributions of the advecting field (i.e. the skewness phenomenon), as it is proposed in Refs. \cite{CheGar19,Muz19}, may allow to reproduce the observed values $H_{\text{\tiny{Eul}}}=1/3$ and $H_{\text{\tiny{Lag}}}=1/2$, possibly on the line $H_{\text{\tiny{Lag}}}=H_{\text{\tiny{Eul}}} + \frac{1}{6}$. 

	\begin{acknowledgments}
		We warmly thank K. Gawedzki for many crucial discussions, M. Adda-Bedia and S. Ciliberto for a critical proofreading of the manuscript. L.C. is partially supported by the Simons Foundation Award ID: 651475. 
	\end{acknowledgments}

%


\clearpage
\onecolumngrid
\appendix

\setcounter{equation}{0}
\setcounter{figure}{0}
\setcounter{table}{0}
\setcounter{page}{1}
\makeatletter
\renewcommand{\theequation}{S\arabic{equation}}
\renewcommand{\thefigure}{S\arabic{figure}}
	\section*{Supplemental Material:\\
	Flow of Spatiotemporal Turbulentlike Random  Fields}
	\begin{center}
		Jason Reneuve$^1$ and Laurent Chevillard$^{2}$\\
		\mbox{~}\\
		\textit{$^1$Laboratoire des Ecoulements G\'eophysiques et Industriels, Universit\'e Grenoble Alpes, CNRS, Grenoble-INP, F-38000 Grenoble, France}\\		
		\textit{$^2$Univ Lyon, Ens de Lyon, Univ Claude Bernard, CNRS, Laboratoire de Physique, 46 all\'ee d'Italie F-69342 Lyon, France}\\

		\vspace{-0.4cm}
	\end{center}

	\subsection*{I. A quick overview of Fractional Gaussian Fields}

We now provide a short  introduction to fractional Gaussian fields (fGfs) on which our random vector field (Eq. (2)) is based. A detailed presentation of these fields is proposed for instance in   \cite{RobVar08,CheRob10,PerGar16,CheGar19,Ren19}. At this stage, to keep the discussion simple, we consider a scalar field $u_a(x)$ in a $d$-dimensional space, i.e. $\boldsymbol{x}\in\mathbb R^d$. We furthermore assume this field Gaussian, statistically homogeneous, isotropic and of zero average, thus fully defined by its covariance function $\mathcal C_a(|\boldsymbol{\ell}|)=\langle u_a(\boldsymbol{x})u_a(\boldsymbol{x}+\boldsymbol{\ell}) \rangle$. Given these assumptions, the Gaussian field $u_a(\boldsymbol{x})$ can be equivalently and conveniently written as the following stochastic integral,
\begin{equation}\label{eq:fGf}
u_a(\boldsymbol{x}) = \int_{\mathbb R^d}g(\boldsymbol{x}-\boldsymbol{y}) W(d^dy),
\end{equation}
where $W$ is a Gaussian white noise of variance $d^dy$, and $g$ a deterministic function (i.e. the filtering kernel) that remains to be determined. This kernel $g$ is related to the covariance $\mathcal C$ as $|\widehat{g}|^2=\widehat{\mathcal C}$, where $\widehat{.}$ stands for the Fourier transform. We choose it such that $u_a$ (i) is a finite-variance process, and (ii) has locally the same regularity as the fractional Brownian motion \cite{ManVan68} of parameter $H_{\text{\tiny{Eul}}}\in]0,1[$. For these reasons, and statistical isotropy, we choose $g$ to be
\begin{equation}
g(\boldsymbol{x}) =\varphi(\boldsymbol{x})||\boldsymbol{x}||_\epsilon^{H_{\text{\tiny{Eul}}}-d/2},
\end{equation}
where $||\boldsymbol{x}||_\epsilon^2 = |\boldsymbol{x}|^2+\epsilon^2$ is a regularized norm over $\epsilon$ which ensures that, at a given $\epsilon>0$, the field $u_a$ is differentiable. The regularizing parameter $\epsilon$ plays the role of the dissipative length scale of turbulence, that goes to 0 as the Reynolds number increases. The cutoff function $\varphi$ is also chosen as an isotropic function of the vector $\boldsymbol{x}$ and allows the introduction of the decorrelation length $L$ (i.e. the integral length scale in the vocabulary of turbulence). As we will see, its precise shape has no impact on the small scale structure of the field $u_a$, besides ensuring that $u_a$ has a finite variance and decorrelates over a given length scale $L$. For these reasons, we choose the isotropic function as $\varphi(\boldsymbol{x})=\varphi(|\boldsymbol{x}|) \propto \exp(-|\boldsymbol{x}|^2/(2L^2))$.

Following the lines developed for instance in \cite{RobVar08,CheRob10,PerGar16,CheGar19,Ren19}, it is straightforward to get for the variance
\begin{align}\label{eq:DefVarFGF}
\lim_{\epsilon\to 0}\langle u_a^2\rangle = \int_{\mathbb R^d}\varphi(\boldsymbol{x}) |\boldsymbol{x}|^{2H_{\text{\tiny{Eul}}}-d}d^dx,
\end{align}
which is finite for any $H_{\text{\tiny{Eul}}}>0$. To investigate the regularity of this field, consider the velocity increment $\delta_{\boldsymbol{\ell}}u_a(\boldsymbol{x})=u_a(\boldsymbol{x}+\boldsymbol{\ell})-u_a(\boldsymbol{x})$ over a given scale $\boldsymbol{\ell}$, and get
\begin{align}\label{eq:DefRegFGF}
\lim_{\epsilon\to 0}\langle (\delta_{\boldsymbol{\ell}}u_a)^2\rangle \build{\sim}_{|\boldsymbol{\ell}|\to 0}^{} \varphi^2(0)c_2|\boldsymbol{\ell}|^{2H_{\text{\tiny{Eul}}}},
\end{align}
where $c_2$ is finite for any $0<H_{\text{\tiny{Eul}}}<1$ and reads, for any unit vector $\boldsymbol{e}$, 
\begin{align*}
c_2 = \int_{\mathbb R^d}\left[|\boldsymbol{x}+\boldsymbol{e}|^{H_{\text{\tiny{Eul}}}-d/2}-|\boldsymbol{x}|^{H_{\text{\tiny{Eul}}}-d/2}\right]^2d^dx.
\end{align*}
The statistical behaviors given in Eqs. \ref{eq:DefVarFGF} and \ref{eq:DefRegFGF} fulfill the constraints of (i) finite variance and (ii) local regularity of parameter $H_{\text{\tiny{Eul}}}\in]0,1[$. Higher-order structure functions are straightforward to get since $u_a$ and its increments are Gaussian. For these reasons, odd-order moments vanish, and even-order ones are given by
$$ \lim_{\epsilon\to 0}\langle (\delta_{\boldsymbol{\ell}}u_a)^{2q}\rangle \build{\sim}_{|\boldsymbol{\ell}|\to 0}^{} \frac{(2q)!}{2^qq!}\varphi^{2q}(0)c_2^q|\boldsymbol{\ell}|^{2qH_{\text{\tiny{Eul}}}},
$$
showing that asymptotically the process $u_a$ is monofractal of parameter $H_{\text{\tiny{Eul}}}$.

	\subsection*{II. Multiplicative chaos as a model of the intermittency phenomenon}

As reviewed in Refs. \cite{RobVar08,CheRob10,PerGar16,CheGar19,Ren19}, a way to incorporate intermittent corrections to fractional Gaussian fields  is to perturb the white noise measure $W$ entering in Eq. (\ref{eq:fGf}) by a positive and independent random weight taken as the exponential of a log-correlated Gaussian field $Y$. Such a procedure requires some care because involved fields are necessarily of infinite variance. In few words,  following a well-posed regularizing procedure, we can give a meaning to the exponential of such a field (see the review \cite{RhoVar14} on mathematical developments of multiplicative chaos theory)  that can be viewed as a continuous, statistically homogeneous and/or stationary version of the discrete cascade models \cite{MenSre87,BenBif93,ArnBac98} used to model intermittency.

Following the lines leading to the Gaussian fractional field $u_a$ (Eq. (\ref{eq:fGf})), we now propose an intermittent version that reads
\begin{equation}\label{eq:fGfMulti}
u_b(\boldsymbol{x}) = \int_{\mathbb R^d}g(\boldsymbol{x}-\boldsymbol{y}) e^{\gamma_{\text{\tiny{Eul}}} Y(y)-\gamma^2_{\text{\tiny{Eul}}} \langle Y^2\rangle} W(d^dy),
\end{equation}
where $Y$ is assumed to be Gaussian, and independent on $W$, and given by 
\begin{equation}\label{eq:defY}
Y(\boldsymbol{y}) =\frac{1}{\sqrt{s_d}}\int_{|\boldsymbol{x}-\boldsymbol{y}|\le L}||\boldsymbol{x}-\boldsymbol{y}||_\epsilon^{-d/2}\widetilde{W}(d^dz),
\end{equation}
with $s_d=2\pi^{d/2}/\Gamma(d/2)$ the surface of the unit sphere in dimension $d$  ($\Gamma$ standing for the usual Gamma function)  and $\widetilde{W}$ an independent white noise measure. The field $Y$ can be seen as a regularized version (over $\epsilon$) of a fGf of vanishing Hurst exponent $H_{\text{\tiny{Eul}}}=0$. It has a vanishing average and its variance can be computed as
$$\langle Y^2\rangle =  \frac{1}{s_d}\int_{|\boldsymbol{z}|\le L} ||\boldsymbol{z}||_\epsilon^{-d}d^dz\build{\sim}_{\epsilon\to 0}^{} \log\frac{1}{\epsilon}.$$
Whereas the variance diverges as $\epsilon\to 0$, its covariance remains bounded over a finite scale $|\boldsymbol{\ell}|$, and we get
$$\lim_{\epsilon\to 0}\langle Y(\boldsymbol{y})Y(\boldsymbol{y}+\boldsymbol{\ell})\rangle =  \frac{1}{s_d}\int_{|\boldsymbol{z}|\le L\wedge |\boldsymbol{z}+\boldsymbol{\ell}|\le L} |\boldsymbol{z}|^{-d/2} |\boldsymbol{z}+\boldsymbol{\ell}|^{-d/2}d^dz\build{\sim}_{|\boldsymbol{\ell}|\to 0}^{} \log\frac{1}{|\boldsymbol{\ell}|}.$$
Because we assumed that the fields $Y$ and $W$ are independent, it is easy to show that the covariance $\mathcal C_b(|\boldsymbol{\ell}|)=\langle u_b(\boldsymbol{x})u_b(\boldsymbol{x}+\boldsymbol{\ell}) \rangle$ is unchanged and equal to the covariance $\mathcal C_a$ of $u_a$. Same conclusions can be drawn for the variance (Eq. (\ref{eq:DefVarFGF})) and second-order structure function (Eq. (\ref{eq:DefRegFGF})). Concerning the behavior at small scales of high-order structure functions, we obtain, for an integer $q\ge 1$, $H_{\text{\tiny{Eul}}}\in]0,1[$ and $\gamma^2_{\text{\tiny{Eul}}}<H_{\text{\tiny{Eul}}}/(q-1)$,
$$ \lim_{\epsilon\to 0}\langle (\delta_{\boldsymbol{\ell}}u_b)^{2q}\rangle \build{\sim}_{|\boldsymbol{\ell}|\to 0}^{} \varphi^{2q}(0)c'_{2q}|\boldsymbol{\ell}|^{2qH_{\text{\tiny{Eul}}}-2q(q-1)\gamma^2_{\text{\tiny{Eul}}}},
$$
where the positive multiplicative factor $c'_{2q}$ can be computed, showing that the process $u_b$ is asymptotically multifractal, its spectrum of exponents being quadratic, of parameter $H_{\text{\tiny{Eul}}}$ and $\gamma^2_{\text{\tiny{Eul}}}$.

	\subsection*{III. Final comments on the structure of the proposed advection field}

The proposed advecting spatiotemporal Eulerian vector field $\boldsymbol{u}(\boldsymbol{x},t) = (u_1,u_2)$ (Eq. (2)) can be seen as a generalization of the scalar field $u_b$ (Eq. (\ref{eq:fGfMulti})) in dimension $d=3$, two dimensions being used for space and one dimension for time. In this case, the area of the unit-sphere is $s_3=4\pi$. The incompressible nature of the vector field $\boldsymbol{u}$ is fulfilled while introducing the vector $\boldsymbol{x}^{\perp}=(-x_2,x_1)$ in the picture, properly normalized such that its has a unit norm as $\epsilon\to 0$.
	
	\clearpage
	
	\subsection*{IV. Alternative propositions and their temporal behavior}
	 
	For the sake of generality, and to make a connection with the propositions of Ref. \cite{ChaGaw03},  let us now consider $d$ dimensions for space $\boldsymbol{x}\in \mathbb R^d$, $t\in \mathbb R$, and vector fields $\boldsymbol{u}=(u_1,\ldots,u_d)\in\mathbb R^d$. Also, to simplify the discussions, let us assume $\boldsymbol{u}$ to be a zero-average Gaussian random vector field, and thus neglect additional intermittent corrections. In this case, assuming furthermore statistical isotropy, homogeneity and stationarity, the vector field $\boldsymbol{u}$ is fully characterized by its correlation function $\mathcal C_{ij}$ that has a rather simple expression in the Fourier space \cite{ChaGaw03}. It reads
	\begin{equation}\label{eq:GeneCorr}
	\mathcal C_{ij}(\boldsymbol{\ell},\tau)=\langle u_i(\boldsymbol{x},t)u_j(\boldsymbol{x}+\boldsymbol{\ell},t+\tau)\rangle= D_2\int \big|\widehat{g}(|\boldsymbol{k}|,\omega)\big|^2 \widehat{P}_{ij}(\boldsymbol{k})e^{2i\pi(\boldsymbol{k}\cdot\boldsymbol{\ell}+\omega\tau)}d^dkd\omega,
	\end{equation}
where $D_2$ is a multiplicative constant taken such that $\langle |\boldsymbol{u}|^2\rangle=\mathcal C_{ii}(\boldsymbol{0},0)=1$ (we adopt Einstein's convention of sum over repeated indices), $\widehat{P}_{ij}(\boldsymbol{k})=\delta_{ij}-\frac{k_ik_j}{|\boldsymbol{k}|^2}$ the Fourier transform of the Leray's projector on divergence free vector fields ($\delta_{ij}$ being the Kronecker symbol), and $\widehat{g}$ a scalar function that depends only on the norm of the wave vector $\boldsymbol{k}$ and frequency $\omega$.

Once the correlation function $\mathcal C_{ij}$ is imposed (Eq. (\ref{eq:GeneCorr})), the corresponding vector field $\boldsymbol{u}(\boldsymbol{x},t)$ can be written as a linear filtering of a Gaussian white noise vector measure $\boldsymbol{W}(d^dx,dt)=\left(W_1(d^dx,dt),\ldots,W_d(d^dx,dt)\right)$, each $W_i$ being independent copies of the $(d+1)$-dimensional scalar white noise, and we note by $\widehat{W}_j(d^dk,d\omega)$ their Fourier transform. This expression reads
\begin{equation}\label{eq:GeneVelField}
	u_i(\boldsymbol{x},t)= \sqrt{D_2}\int \big|\widehat{g}(|\boldsymbol{k}|,\omega)\big| \widehat{P}_{ij}(\boldsymbol{k})e^{2i\pi(\boldsymbol{k}\cdot\boldsymbol{x}+\omega t)}\widehat{W}_j(d^dk,d\omega).
	\end{equation}

	\subsubsection*{IV.a. Considerations on three different random vector fields}
	
	Let us now study three different incompressible random vector fields, call them $\boldsymbol{u}^a$, $\boldsymbol{u}^b$ and $\boldsymbol{u}^c$, whose spatiotemporal structure is governed by the kernel $\widehat{g}(|\boldsymbol{k}|,\omega)$ entering in Eq. (\ref{eq:GeneVelField}). 
	
	We first consider a kernel leading to similar behaviors as the field $\boldsymbol{u}$ used in the first part of this Letter (Eq. (2)), a situation for which, roughly speaking, space and time are treated indifferently. Such a kernel would read,	
	\begin{equation}\label{eq:FirstG}
	 \big|\widehat{g}(|\boldsymbol{k}|,\omega)\big|^2 = \frac{e^{-4\pi\epsilon \sqrt{|\boldsymbol{k}|^2 + \omega^2}}}{\left[D_3^2\left(|\boldsymbol{k}|^2 + L^{-2} \right)+\omega^2\right]^{\frac{d+1}{2}+H_{\text{\tiny{Eul}}}}}
	\end{equation}
where $D_3$ is a constant that has dimension of a velocity (i.e. a length over a time). There, $\epsilon$ and $L$ play the same roles as in Eq. (2), corresponding to respectively a small scale regularization ensuring differentiability and a large scale cut-off that warrants a finite variance. We discard any further multiplicative factor that is eventually included in the constant $D_2$  such that the corresponding velocity field is of unit variance (Eq. (\ref{eq:GeneVelField})). When $D_3=1$ and $d=2$, the main difference between the field given in Eq. (2) and the one governed by Eq. (\ref{eq:FirstG}) originates from these small and large scale regularizations, and we expect very similar behaviors at small scales as those considered in Figs. 2 and 3 when $\epsilon\rightarrow 0$.

As it is considered in Ref. \cite{ChaGaw03}, let us now consider a  kernel that treats time and space differently. The proposition of Ref. \cite{ChaGaw03} assumes an exponential correlation in time of characteristic duration given by a power-law of the wave number $|\boldsymbol{k}|$. Equivalently, it reads in the wave vector and frequency domains
	\begin{equation}\label{eq:SecondG}
	 \big|\widehat{g}(|\boldsymbol{k}|,\omega)\big|^2 =  \frac{\left(|\boldsymbol{k}|^2 + L^{-2} \right)^{\beta}}{D_3^2\left(|\boldsymbol{k}|^2 + L^{-2} \right)^{2\beta}+\omega^2}\frac{e^{-4\pi\epsilon \sqrt{|\boldsymbol{k}|^2 + \omega^2}}}{\left(|\boldsymbol{k}|^2 + L^{-2} \right)^{\frac{d}{2}+H_{\text{\tiny{Eul}}}}},
	\end{equation}
where now $D_3$ has dimension of a length to the power $2\beta$ over time, as argued in  Ref. \cite{ChaGaw03}. We can recognize in Eq. (\ref{eq:SecondG}) a Lorentzian term, reminiscent of an exponential correlation in time. As we will see, the free parameter $\beta$ entering in Eq. (\ref{eq:SecondG}) governs the temporal structure of the field.

\begin{table}[ht]
\begin{center}
 \begin{tabular}{||c| c| c| c | c | c ||} 
 \hline
Field  & Kernel  $\big|\widehat{g}(|\boldsymbol{k}|,\omega)\big|^2$ & Spatial $\left\langle \big|\delta_{\boldsymbol{\ell}} \boldsymbol{u}\big|^2\right\rangle$ & Temporal $\left\langle \big|\delta_{\tau} \boldsymbol{u}\big|^2\right\rangle$ &  $\tau_c(|\boldsymbol{\ell}|)$ (Eq. (\ref{eq:TauCEll}))&  $\tau_e(|\boldsymbol{\ell}|)$ (Eq. (\ref{eq:TauEEll}))\\ [0.5ex] 
 \hline\hline
$\boldsymbol{u}^a$  & Eq. (\ref{eq:FirstG})  & $ |\boldsymbol{\ell}|^{2H_{\text{\tiny{Eul}}}}$  & $\tau ^{2H_{\text{\tiny{Eul}}}}$  & $|\boldsymbol{\ell}|$ & $|\boldsymbol{\ell}|^{1-H_{\text{\tiny{Eul}}}}$ \\ 
 \hline
$\boldsymbol{u}^b$  & Eq. (\ref{eq:SecondG}) with $\beta=\frac{1}{2}$ & $ |\boldsymbol{\ell}|^{2H_{\text{\tiny{Eul}}}}$  & $ \tau ^{2H_{\text{\tiny{Eul}}}}$ & $\build{}_{|\boldsymbol{\ell}|^{2(1-H_{\text{\tiny{Eul}}})} \text{ for } H_{\text{\tiny{Eul}}}\ge 1/2}^{|\boldsymbol{\ell}| \text{ for } H_{\text{\tiny{Eul}}}\le 1/2 }$ & $ |\boldsymbol{\ell}|^{1-H_{\text{\tiny{Eul}}}}$ \\
 \hline
$\boldsymbol{u}^c$  &  Eq. (\ref{eq:SecondG}) with $\beta=\frac{1-H_{\text{\tiny{Eul}}}}{2}$ & $ |\boldsymbol{\ell}|^{2H_{\text{\tiny{Eul}}}}$  & $\build{}_{\tau^{2} \text{ for } H_{\text{\tiny{Eul}}}\ge 1/2}^{\tau ^{\frac{2H_{\text{\tiny{Eul}}}}{1-H_{\text{\tiny{Eul}}}}}\text{ for } H_{\text{\tiny{Eul}}}\le 1/2 }$ & $|\boldsymbol{\ell}|^{1-H_{\text{\tiny{Eul}}}}$  & $|\boldsymbol{\ell}|^{1-H_{\text{\tiny{Eul}}}}$ \\[1ex] 
 \hline
\end{tabular}
\end{center}
\caption{Definition of the three incompressible random fields $\boldsymbol{u}^a$, $\boldsymbol{u}^b$ and $\boldsymbol{u}^c$, based on  Eq. (\ref{eq:GeneVelField}). Case $a$: we use the kernel $\big|\widehat{g}(|\boldsymbol{k}|,\omega)\big|$ whose square is provided in Eq. (\ref{eq:FirstG}). Case $b$ (resp. $c$), we use the kernel given in Eq. (\ref{eq:SecondG}) with $\beta=\frac{1}{2}$ (resp. $\beta=\frac{1-H_{\text{\tiny{Eul}}}}{2}$). In all cases, we consider any space dimension $d\ge 2$ and $H_{\text{\tiny{Eul}}}\in ]0,1[$. We then provide the behaviors of the second moment of the spatial $\delta_{\boldsymbol{\ell}} \boldsymbol{u}(\boldsymbol{x},t)=\boldsymbol{u}(\boldsymbol{x}+\boldsymbol{\ell},t)-\boldsymbol{u}(\boldsymbol{x},t)$ and temporal $\delta_{\tau} \boldsymbol{u}(\boldsymbol{x},t)=\boldsymbol{u}(\boldsymbol{x},t+\tau)-\boldsymbol{u}(\boldsymbol{x},t)$ velocity increments. All behaviors are understood taking first the limit $\epsilon\to 0$, and only then the respective length $|\boldsymbol{\ell}|$ or temporal $\tau$ scales going to zero. Similarly, in the same double limit, we give the scale dependence of the two characteristic durations $\tau_c(|\boldsymbol{\ell}|)$ (Eq. (\ref{eq:TauCEll})) and $\tau_e(|\boldsymbol{\ell}|)$ (Eq. (\ref{eq:TauEEll})).}\label{tab:DefABC}
\end{table}

As it is proposed in Ref. \cite{ChaGaw03}, in order to illustrate the temporal behaviors of the random velocity fields induced by the kernels of Eqs. \ref{eq:FirstG} and \ref{eq:SecondG}, we consider the correlation time $\tau_c(|\boldsymbol{\ell}|)$ of the velocity differences $\delta_{\boldsymbol{\ell}} \boldsymbol{u}(\boldsymbol{x},t)=\boldsymbol{u}(\boldsymbol{x}+\boldsymbol{\ell},t)-\boldsymbol{u}(\boldsymbol{x},t)$, defined by
	\begin{equation}\label{eq:TauCEll}
	 \tau_c(|\boldsymbol{\ell}|) = \frac{1}{\left\langle \big|\delta_{\boldsymbol{\ell}} \boldsymbol{u}\big|^2\right\rangle}\int_0^{\infty}\big\langle \delta_{\boldsymbol{\ell}} \boldsymbol{u}(\boldsymbol{x},t)\cdot \delta_{\boldsymbol{\ell}} \boldsymbol{u}(\boldsymbol{x},t+\tau)\big\rangle d\tau
	\end{equation}
and the eddy turnover timescale $\tau_e(|\boldsymbol{\ell}|)$ defined by
	\begin{equation}\label{eq:TauEEll}
	 \tau_e(|\boldsymbol{\ell}|) = \frac{|\boldsymbol{\ell}|}{\sqrt{\left\langle \big|\delta_{\boldsymbol{\ell}} \boldsymbol{u}\big|^2\right\rangle}}.
	\end{equation}

We define in table \ref{tab:DefABC} three incompressible random fields $\boldsymbol{u}^a$, $\boldsymbol{u}^b$ and $\boldsymbol{u}^c$ with differing spatiotemporal structures, depending on the choice of the kernel (Eq. (\ref{eq:FirstG}) or Eq. (\ref{eq:SecondG})). Whereas $\boldsymbol{u}^a$ is very similar to the one we have defined in the core of this Letter (Eq. (2)), the main differences lying in the methods of regularization at small (over $\epsilon$) and large (over $L$) length scales, the temporal structures of $\boldsymbol{u}^b$ and $\boldsymbol{u}^c$ are of different nature. In all cases, the spatial structure is similar, as it can be seen from the behavior at small scales of the spatial velocity increment (Third column of table \ref{tab:DefABC}): the parameter $H_{\text{\tiny{Eul}}}$ governs completely the regularity in space. The fields $\boldsymbol{u}^a$ and $\boldsymbol{u}^b$ share also a similar temporal regularity, as evidenced by the behavior at small timescales of the temporal velocity increment (Fourth column of table \ref{tab:DefABC}), the same parameter $H_{\text{\tiny{Eul}}}$ characterizing the temporal regularity. Also, the time correlation $\tau_c(|\boldsymbol{\ell}|)$ (Eq. (\ref{eq:TauCEll})) of the velocity increments over $\boldsymbol{\ell}$ (Fifth column of table \ref{tab:DefABC}) is always much smaller than the eddy turnover timescale $\tau_e(|\boldsymbol{\ell}|)$ (Eq. (\ref{eq:TauEEll})). Let us notice that $\tau_c$ for the field $\boldsymbol{u}^b$ undergoes a transition when $H_{\text{\tiny{Eul}}}=1/2$ that is related to the existence of the integral entering in Eq. (\ref{eq:TauCEll}), without changing the fact that $\tau_c\ll \tau_e$ as $\boldsymbol{\ell}\to 0$. Note also that having $\tau_c$ proportional to $|\boldsymbol{\ell}|$ is characteristic of the sweeping of the small scales by the large scales. The time regularity of $\boldsymbol{u}^c$ is rather different from the one of the two other fields, and is always smoother whatever the value of $H_{\text{\tiny{Eul}}}$. Furthermore, $\tau_c$ is of the same order as $\tau_e$, as it is discussed in Ref. \cite{ChaGaw03}, and is, in this sense, not affected by the sweeping by the large scales. Again, note a transition in the behavior of the temporal increments as $H_{\text{\tiny{Eul}}}$ crosses $1/2$, which is again due to existence of some underlying integrals.

	\subsubsection*{IV.b. Numerical simulations}

\begin{figure}
		\begin{center}
			\includegraphics[width=.8\columnwidth]{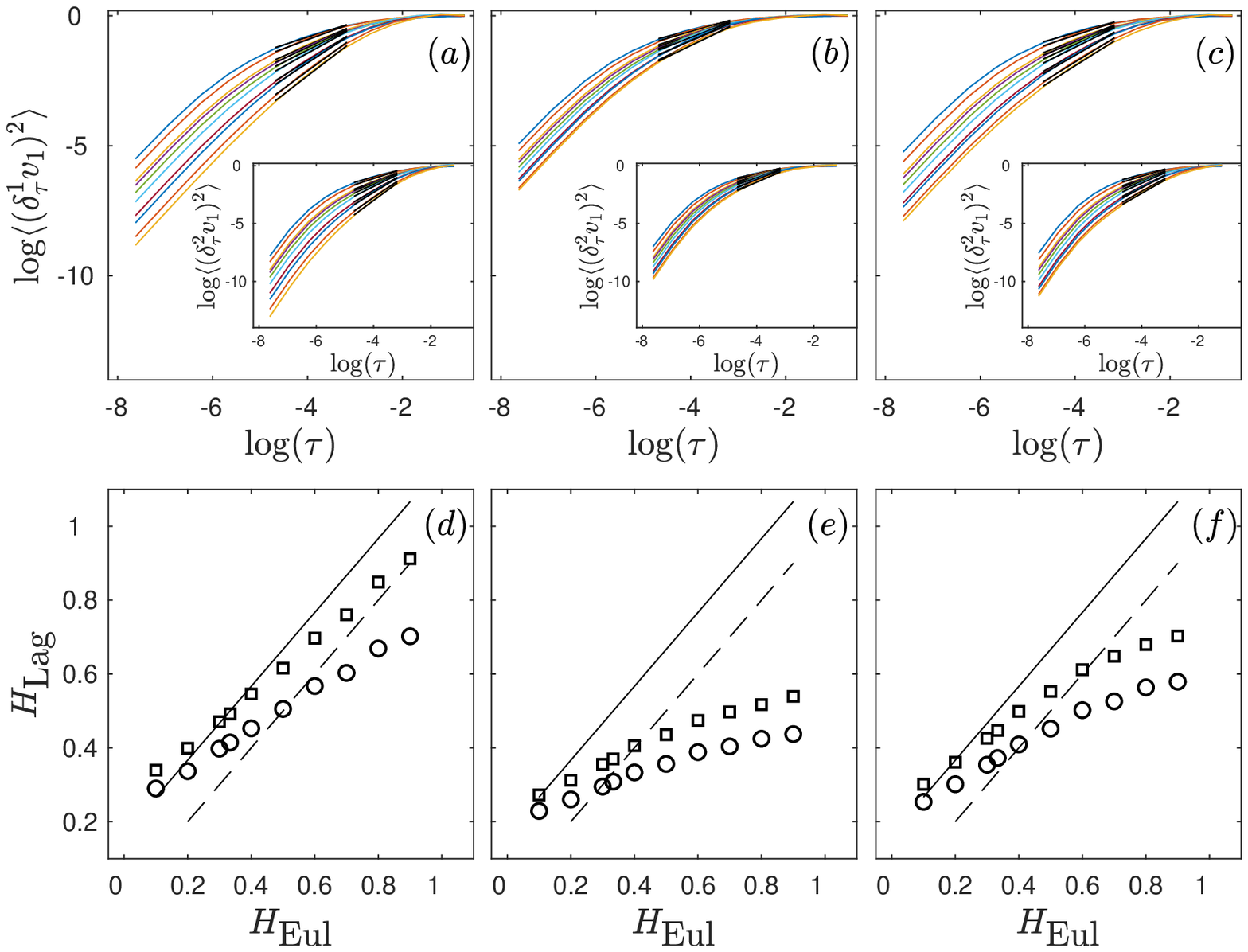}
			\vspace{-0.2cm}
			\caption{Similar plot as in Fig. 2, but for the fields $\boldsymbol{u}^a$ ((a) and (d)), $\boldsymbol{u}^b$ ((b) and (e)) and $\boldsymbol{u}^c$ ((c) and (f)). Fields are defined in Table \ref{tab:DefABC}. } 
			\label{fig:ABCVF}				
		\end{center}
		\vspace{-0.7cm}
	\end{figure}

We perform numerical simulations of the fields $\boldsymbol{u}^a$, $\boldsymbol{u}^b$ and $\boldsymbol{u}^c$ for $d=2$, and extract their respective induced Lagrangian trajectories, in a very similar manner as in the beginning of the article. We use $N=2^{11}$ collocation points in each spatial or temporal direction, using $dx=dt=1/N$, $\epsilon=2dx$, $L=1/2$ and $D_3=1$. Recall that $D_2$ is chosen such that $\langle|\boldsymbol{u}^{a,b,c}|^2\rangle=1$. Again, we track $2^{14}$ particles, initially uniformly distributed in the unit square, and display the results of our statistical analysis of velocity along of trajectories in Fig. \ref{fig:ABCVF}.

Indeed, as expected and mentioned before, statistical properties of Lagrangian velocity extracted from $\boldsymbol{u}^a$ are displayed in Figs.  \ref{fig:ABCVF}(a) and (d) and are found to be very similar to those extracted from $\boldsymbol{u}$ (Eq. (2)), although the proportionality of $H_{\text{\tiny{Lag}}}$ to $H_{\text{\tiny{Eul}}}$ is not as clear as in Fig. 2. This is very probably due to the different choices that have been made to define regularizations over small and large length scales. Again, and we will come back to this point later in this Appendix, we obtain very similar behaviors if we consider a frozen-in-time version of $\boldsymbol{u}^a$ to advect the particles (data not shown).

The behavior of particles in the fields $\boldsymbol{u}^b$ and $\boldsymbol{u}^c$ is found to be different from those seen in $\boldsymbol{u}^a$. Concerning trajectories extracted from $\boldsymbol{u}^b$ (resp. $\boldsymbol{u}^c$), we show the statistical properties of Lagrangian velocity in  Figs.  \ref{fig:ABCVF}(b) and (e) (resp. Figs.  \ref{fig:ABCVF}(c) and (f)). As it can be seen in Figs.  \ref{fig:ABCVF}(e) and (f), $H_{\text{\tiny{Lag}}}$ does not behave linearly with $H_{\text{\tiny{Eul}}}$, even when estimated with the second-order velocity increment. We nonetheless see that trajectories of $\boldsymbol{u}^c$ are consistent with $H_{\text{\tiny{Lag}}}=H_{\text{\tiny{Eul}}}+1/6$  when $H_{\text{\tiny{Eul}}}\le 0.4$, which includes in particular the Kolmogorov's values $H_{\text{\tiny{Eul}}}=1/3$ and $H_{\text{\tiny{Lag}}}=1/2$. For larger values of $H_{\text{\tiny{Eul}}}$, say $H_{\text{\tiny{Eul}}}\ge 1/2$, we evidence a much weaker dependence of $H_{\text{\tiny{Lag}}}$ on $H_{\text{\tiny{Eul}}}$. As far as $\boldsymbol{u}^b$  is concerned, this phenomenon is very probably due to the transition undergone by the correlation time  $\tau_c$ (Eq. (\ref{eq:TauCEll})) of increments over $\boldsymbol{\ell}$, as it is stated in Table \ref{tab:DefABC}. Similarly, concerning $\boldsymbol{u}^c$, we interpret the weakening of the dependence of $H_{\text{\tiny{Lag}}}$ on $H_{\text{\tiny{Eul}}}$ by the transition undertaken by the temporal velocity increment $\left\langle \big|\delta_{\tau} \boldsymbol{u}\big|^2\right\rangle$ at $H_{\text{\tiny{Eul}}}=1/2$, as it is recalled in Table \ref{tab:DefABC}. We also estimated the statistical behaviors of Lagrangian velocity in frozen-in-time versions of the advecting fields $\boldsymbol{u}^b$ and $\boldsymbol{u}^c$, and found behaviors similar to the ones observed for $\boldsymbol{u}^a$ (data not shown). Hence, frozen versions of $\boldsymbol{u}^b$ and $\boldsymbol{u}^c$  lead to Lagrangian velocities that are different from those obtained from time-evolving ones, and this makes a clear difference with fields $\boldsymbol{u}$ (Eq. (2)) and $\boldsymbol{u}^a$.

	\subsubsection*{IV.c. Additional numerical simulations at high resolutions of frozen-in-time advecting fields}

			\begin{figure}
		\begin{center}
			\includegraphics[width=.6\columnwidth]{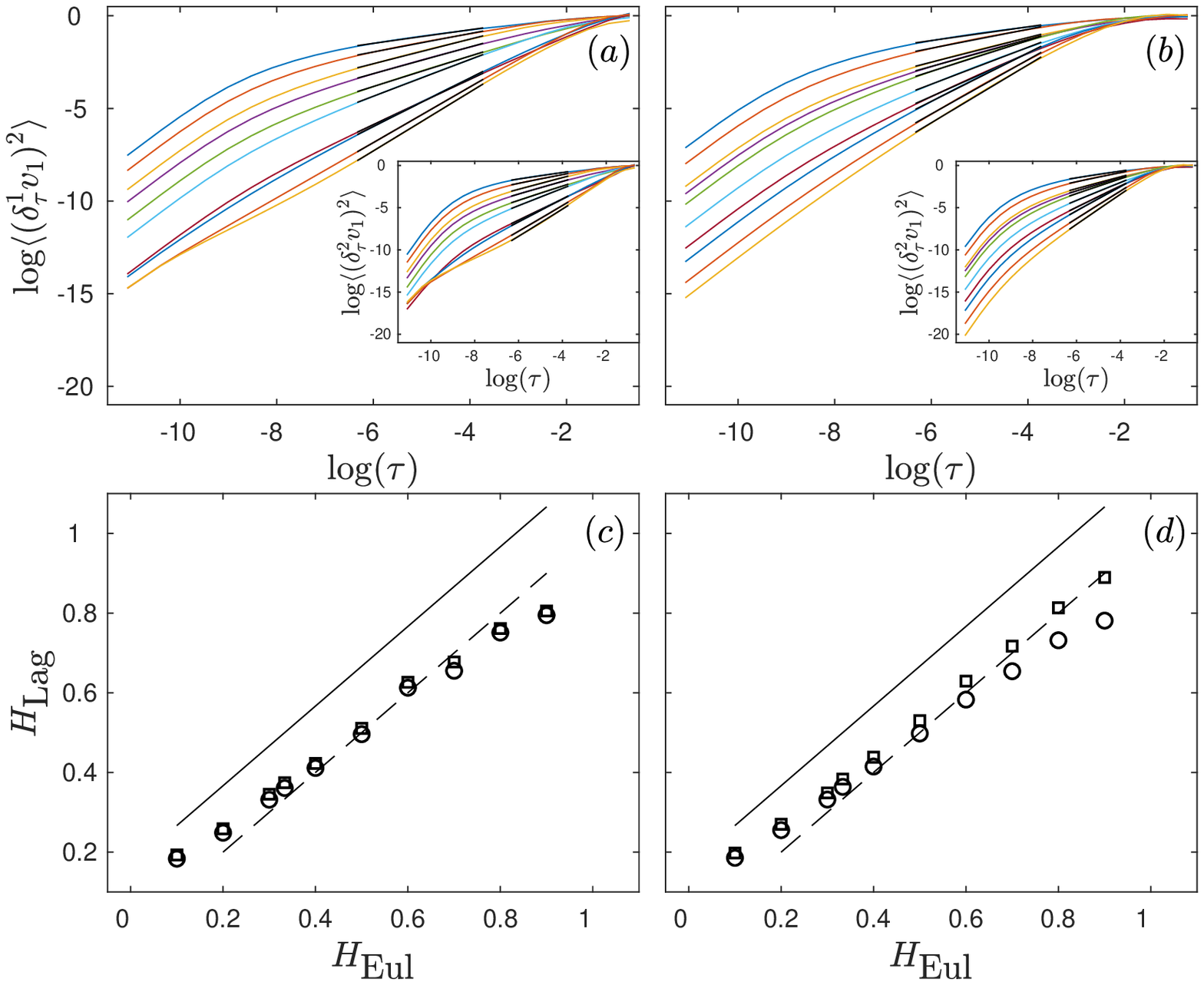}
			\vspace{-0.2cm}
			\caption{Similar plot as in Fig. 2, but in (a) and (c) for the field $\boldsymbol{u}^{\text{\tiny{2D}}}$ (Eq. (\ref{eq:VelField2D})) and in (b) and (d) for $\boldsymbol{u}^{\text{\tiny{2D}},abc}$ (Eq. (\ref{eq:GeneVelField2D})) using $N=2^{16}$ collocation points in each spatial direction.} 
			\label{fig:Frozen2D}				
		\end{center}
		\vspace{-0.7cm}
	\end{figure}

We have seen that statistics of Lagrangian trajectories are very similar for the fields $\boldsymbol{u}$ (Eq. (2)) and $\boldsymbol{u}^a$, as it can be seen in Figs. 2(c) and  \ref{fig:ABCVF}(d). As we have shown in Fig. 2(d), very similar behaviors are observed for a frozen version of $\boldsymbol{u}$. In the same way, a frozen version of $\boldsymbol{u}^a$ also gives similar results (data not shown). The fact that statistics of Lagrangian velocity are the same in evolving and frozen-in-time versions of these fields is consistent with the treatment on the same foot of space and time in $\boldsymbol{u}$ and $\boldsymbol{u}^a$. Recall also that frozen versions of $\boldsymbol{u}^b$ and $\boldsymbol{u}^c$ give also similar behaviors to those obtained with evolving or frozen versions of $\boldsymbol{u}$ and $\boldsymbol{u}^a$ (data not shown), whereas time evolving versions of  $\boldsymbol{u}^b$ and $\boldsymbol{u}^c$ give different behaviors, as shown in Figs. \ref{fig:ABCVF}(e) and (f). In particular, we evidence a transition when $H_{\text{\tiny{Eul}}}$ crosses $1/2$. Again, it is expected since time is taken into account in a different way than space, as it can be seen in the functional form of their kernel (Eq. (\ref{eq:SecondG})).

We report the results of highly resolved simulations of $d=2$ dimensional, purely spatial, advecting fields, to clarify whether the observed Lagrangian Hurst exponent $H_{\text{\tiny{Lag}}}$ is closer to its Eulerian counterpart  $H_{\text{\tiny{Eul}}}$ or to $H_{\text{\tiny{Eul}}}+\frac{1}{6}$. Performing simulations at a much higher resolution, as we will eventually do, would allow us to decipher between genuine inertial range scaling behaviors (for $\epsilon\ll \tau\ll L$) and curvature effects related to regularizations at small $\epsilon$ and large $L$ scales. 

Assuming $d=2$ and discarding the temporal dimension, the advecting velocity field $\boldsymbol{u}$ (Eq. (2)) reduces to 
	\begin{equation}\label{eq:VelField2D}
		\boldsymbol{u}^{\text{\tiny{2D}}}(\boldsymbol{x}) =  \sqrt{D_2}\int_{\boldsymbol{y}\in\mathbb R^2} \varphi(\boldsymbol{x}-\boldsymbol{y})\frac{(\boldsymbol{x}-\boldsymbol{y})^\perp}{||\boldsymbol{x}-\boldsymbol{y},0||_\epsilon}||\boldsymbol{x}-\boldsymbol{y},0||_\epsilon^{H_{\text{\tiny{Eul}}}-1} W(d^2y),
	\end{equation}
and Eq. (\ref{eq:GeneVelField}) reduces to
\begin{equation}\label{eq:GeneVelField2D}
	\boldsymbol{u}^{\text{\tiny{2D}},abc}(\boldsymbol{x})= \sqrt{D_2}\int  \frac{e^{-2\pi\epsilon |\boldsymbol{k}|}}{\sqrt{|\boldsymbol{k}|^2 + L^{-2}}^{1+H_{\text{\tiny{Eul}}}}} \widehat{\boldsymbol{P}}(\boldsymbol{k})e^{2i\pi\boldsymbol{k}\cdot\boldsymbol{x}}\widehat{\boldsymbol{W}}(d^2k),
	\end{equation}
where $(\boldsymbol{P})_{ij}=P_{ij}$ is acting on the white noise vector $\boldsymbol{W}$. Note that the $2D$ versions of $\boldsymbol{u}^a$, $\boldsymbol{u}^b$ and $\boldsymbol{u}^c$ coincide with $\boldsymbol{u}^{\text{\tiny{2D}},abc}$ (Eq. (\ref{eq:GeneVelField2D})). A quick look at the expressions provided in Eqs. \ref{eq:VelField2D} and \ref{eq:GeneVelField2D} confirms that $\boldsymbol{u}^{\text{\tiny{2D}}}$ and $\boldsymbol{u}^{\text{\tiny{2D}},abc}$ share similar statistics, besides the regularization procedures at small and large length scales. Once again, the multiplicative constant $D_2$ is defined such that respective fields are of unit-variance.

We perform numerical simulations of the fields $\boldsymbol{u}^{\text{\tiny{2D}}}$ and $\boldsymbol{u}^{\text{\tiny{2D}},abc}$, using for $\epsilon$ and $L$ the values given respectively for $\boldsymbol{u}$ (Eq. (2)) and in the former section, and extract from them Lagrangian velocity. Working with only bidimensional versions of these fields allows to use $N=2^{16}$ collocation points in each spatial direction. We display the results of the statistical analysis of Lagrangian velocity in Fig. \ref{fig:Frozen2D}.

As expected, we obtain statistics that are very similar for the two fields, and clearly observe a Lagrangian Hurst exponent $H_{\text{\tiny{Lag}}}$ that is closer to  $H_{\text{\tiny{Eul}}}$ than to $H_{\text{\tiny{Eul}}}+\frac{1}{6}$. Moreover, differences between estimations of $H_{\text{\tiny{Lag}}}$ based on first and second velocity increments disappear. Thus, if indeed frozen-in-time advecting fields give the same picture as evolving fields $\boldsymbol{u}$ and $\boldsymbol{u}^a$, then we can conclude that observations made in Fig. 2 are subject to the influence of cut-off methods at small and large length scales, and that the Lagrangian Hurst exponent is given in a good approximation by $H_{\text{\tiny{Lag}}}\approx H_{\text{\tiny{Eul}}}$.

			\begin{figure}
		\begin{center}
			\includegraphics[width=.6\columnwidth]{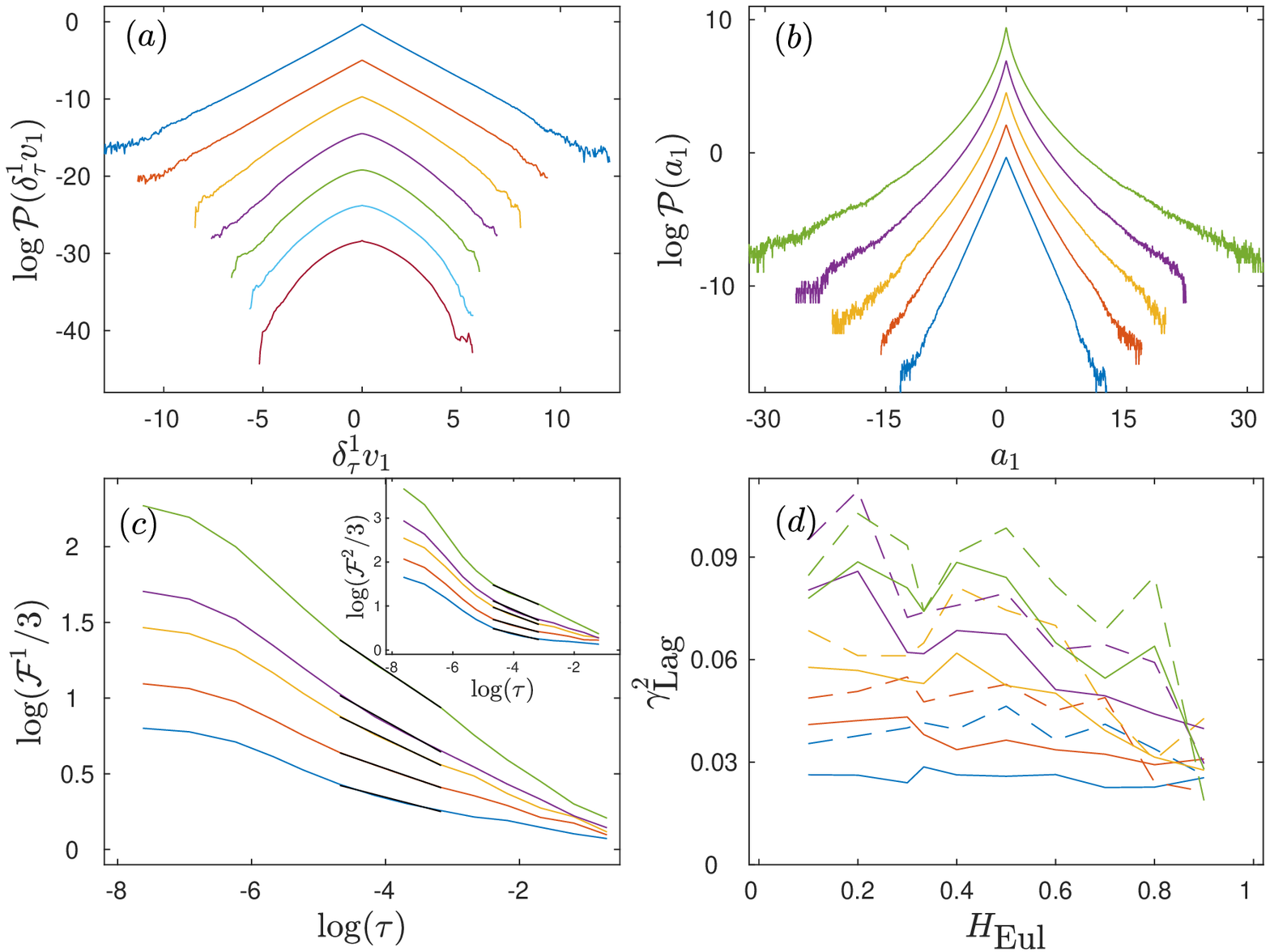}
			\vspace{-0.2cm}
			\caption{Similar plot as in Fig. 3, but for the frozen-in-time two-dimensional velocity field $\boldsymbol{u}^{\text{\tiny{2D}}}$ (Eq. (\ref{eq:VelField2DGamma})), using again $N=2^{11}$, $\epsilon=4dx$, $L=1/2$, over thirty instances of $\boldsymbol{u}^{\text{\tiny{2D}}}$ (instead of ten).}  
			\label{fig:EvolvGammaFrozen2D}				
		\end{center}
		\vspace{-0.7cm}
	\end{figure}

	\subsubsection*{IV.d. Implied intermittent corrections on Lagrangian velocity by frozen-in-time advecting fields}

We finally report the results of a last numerical study, in order to check whether the frozen-in-time advecting fields induce similar intermittent corrections on Lagrangian velocities, as it was observed in Fig. 3 for time-evolving fields. To do so, we use a purely bi-dimensional frozen-in-time velocity field $\boldsymbol{u}^{\text{\tiny{2D}}}$ as in Eq. (\ref{eq:VelField2D}) to advect particles, with possible additional intermittent corrections, given by
	\begin{equation}\label{eq:VelField2DGamma}
		\boldsymbol{u}^{\text{\tiny{2D}}}(\boldsymbol{x}) =  \sqrt{D_2}\int_{\boldsymbol{y}\in\mathbb R^2} \varphi(\boldsymbol{x}-\boldsymbol{y})\frac{(\boldsymbol{x}-\boldsymbol{y})^\perp}{||\boldsymbol{x}-\boldsymbol{y},0||_\epsilon}||\boldsymbol{x}-\boldsymbol{y},0||_\epsilon^{H_{\text{\tiny{Eul}}}-1} e^{\gamma_{\text{\tiny{Eul}}} Y^{\text{\tiny{2D}}}(y)-\gamma^2_{\text{\tiny{Eul}}} \langle (Y^{\text{\tiny{2D}}})^2\rangle}W(d^2y),
	\end{equation}
where the expression of the scalar Gaussian field $Y^{\text{\tiny{2D}}}$ is provided in Eq. (\ref{eq:defY}), assumed independent of the underlying white noise $W$, with $d=2$ and $s_2=2\pi$. We then reproduce the results of the statistical analysis of induced Lagrangian trajectories in Fig. \ref{fig:EvolvGammaFrozen2D}, in a similar way to what was done in Fig. 3. We confirm, as it is displayed in Fig. \ref{fig:EvolvGammaFrozen2D}(a) that Lagrangian velocity obtained from a Gaussian version of the advecting field, i.e. using Eq. (\ref{eq:VelField2D}) or equivalently Eq. (\ref{eq:VelField2DGamma}) with $\gamma_{\text{\tiny{Eul}}}=0$, exhibits intermittent and non-Gaussian features. Also, as $\gamma_{\text{\tiny{Eul}}}$ increases, we observe larger and larger intermittent corrections in the Lagrangian framework, in a very similar quantitative manner as for an advecting time-evolving field (see Fig. 3). Note that, even when using 30 independent instances of the advecting field instead of 10, the statistical convergence of high order moments of Lagrangian velocity increments is not guaranteed, especially for the highest values of $\gamma_{\text{\tiny{Eul}}}$. This can be broadly understood by considering the lack of mixing induced by frozen fields as noticed in Fig. 1, and more generally the lesser randomness that is injected into the system, compared to time-evolving fields. 

Although intermittent corrections on Lagrangian velocity of static and evolving fields appear very similar, let us underline some differences concerning the estimated dependence of $\gamma_{\text{\tiny{Lag}}}$ on $H_{\text{\tiny{Eul}}}$. Comparing the results displayed in figures 3(d) and \ref{fig:EvolvGammaFrozen2D}(d), we can see a tendency of $\gamma_{\text{\tiny{Lag}}}$ to increase with $H_{\text{\tiny{Eul}}}$ for the time-evolving case, whereas an opposite behavior is evidenced in the static one. One should take these overall trends with caution since, as explained, the statistical convergence of fourth-order statistics could be questioned. Nonetheless, it would be very interesting and ambitious to understand the influence of the temporal structure of the advecting fields on Lagrangian intermittency from a theoretical point of view.  

Finally, note that the behavior of the Lagrangian flatness in our model also mimics in a realistic manner some aspects of the differential action of viscosity at finite Reynolds number. On figures 3(c) and \ref{fig:EvolvGammaFrozen2D}(c), we can observe the rapid increase of flatness in the near-dissipative range, a finite-Reynolds effect that is typically observed experimentally and in DNS \cite{CheRou03,ArnBen08}.

\end{document}